# The energy balancing parameter

WALTON R. GUTIERREZ
*waltong@touro.edu*

A parameter method is introduced in order to estimate the relationship among the various variables of a system in equilibrium, where the potential energy functions are incompletely known or the quantum mechanical calculations very difficult. No formal proof of the method is given; instead, a sufficient number of valuable examples are shown to make the case for the method's usefulness in classical and quantum systems. The mathematical methods required are quite elementary: basic algebra and minimization of power functions. This method blends advantageously with a simple but powerful approximate method for quantum mechanics, sidestepping entirely formal operators and differential equations. It is applied to the derivation of various well-known results involving centrally symmetric potentials for a quantum particle such as the hydrogen-like atom, the elastic potential and other cases of interest. The same formulas provide estimates for previously unsolved cases.

## I. INTRODUCTION

In many actual physical systems, the relevant potential energy functions are incompletely known, for example, the potential energy functions of protons and neutrons in the nuclei of atoms and the repulsive potential of the ions in crystalline ionic solids (e.g. salts). Most models of stable equilibrium include an attractive energy potential, a possible repulsive energy potential, and kinetic energy specifications. The kinetic energy may be small or not, may be classical or quantum mechanically determined. It is remarkable that position-dependent energy potentials are among the most durable energy concepts, remaining intact even in a quantum system like the well-known electrostatic potential and the elastic potential.
To gain an insight into the variables relevant to the potential energy functions of a system in equilibrium, a parameter method is introduced. Values of this parameter capture the equilibrium point between the attractive and repulsive potential energy functions. The resulting relations, when supplemented with adequate experimental data, can determine the possible values of the parameter. Since only very basic calculus and algebra operations are required, the method is easy to apply and therefore can serve as a simplified yet productive introduction to atomic and quantum physics.
The 2$^{nd}$ section introduces the method with a straightforward example. In the 3$^{rd}$ section, a way to incorporate approximate quantum mechanical calculations is elaborated around ideas of Louis de Broglie, without using the wave equation of Erwin Schrödinger. The last section shows the parameter method applied to crystalline ionic solids. An application of this method to the nucleus is presented in a separate publication[1].

## II. PARTICLE IN A POTENTIAL WELL

Consider a particle with charge q that is in the electric field of a charge Q of the same sign. The repulsive electrostatic potential is $k_C qQ/r$ with $qQ > 0$ and $k_C$ = Coulomb's constant. The charge is held by an attractive linear potential $k_1 r$, $k_1$ = constant > 0, where r is the distance of the particle to a center of coordinates. The total potential energy of q is

$$V(r) = k_1 r + k_C qQ/r. \tag{1}$$

In this example, the potential energy is fully known, so we can use it to verify the validity of the energy balancing parameter method introduced below.
The classical static equilibrium is obtained by finding the minimum of the potential energy function. Differentiating the energy function

$$\partial V(r)/\partial r = k_1 - k_C qQ/r^2 = 0, \tag{2}$$

we find the following solution for the equilibrium point

$$r = (k_C qQ/k_1)^{1/2} \equiv r_0, \tag{3}$$

and the potential energy value at this point is

$$E_0 = V(r_0) = 2k_C qQ/r_0 = 2(k_1 k_C qQ)^{1/2}. \tag{4}$$

If this potential is part of a quantum mechanical system, the particle does not reach the lowest value of the potential; instead, it settles at the ground state energy value $E_1$ that is above $E_0$

$$E_1 = E_0 + a_1 h\nu, \tag{5}$$

where h = Plank's constant, $a_1$ = dimensionless constant > 0, $\nu$ = a frequency.
See Fig. 1 for an illustration of the potential and the energies $E_1$, $E_0$. The exact value of $E_1$ = quantum ground state energy, has to be determined by solving the Schrödinger equation of that system, or by the application of other quantum mechanical procedures. However in this presentation, the exact value of $E_1$ is not needed. For a reference on the Schrödinger equation, see any textbook on standard quantum mechanics or modern physics for an introductory presentation[2].
There is an interesting explanation for the requirement that $E_1 > E_0$. If we assume $E_1 = E_0$, then the velocity is zero at a position $r = r_0$. But this is not possible for a quantum particle because these values contradict the principle of uncertainty. Stated in simplified terms, this principle requires that the velocity and position along a specific direction can not be exactly and simultaneously known for atomic and subatomic size objects.
The energy balancing parameter method is now applied to the above example. The basic general hypothesis of the method is that the equilibrium of the system is found for a value (or values) of a dimensionless parameter $p_0$ expressing the balance of the attractive and repulsive potential. The equilibrium of the system is now established by the condition that the attractive potential is proportional to the repulsive potential. The proportion is set with the energy balancing parameter $p_0$ in the following way,

$$k_1 r = p_0 k_C qQ/r, \tag{6}$$

which has the solution

$$r = (p_0 k_C qQ/k_1)^{1/2} = r_p. \tag{7}$$

This last solution includes the classical solution (3) with $p_0 = 1$. Replacing (7) into (1), we obtain $V(r_p) = E_0(1 + p_0)/2p_0^{1/2}$; the ground state energy $E_1$ is related to the parameter by requiring that $V(r_p) = E_1$, and combining with the previous equation, $E_1/E_0 = (1 + p_0)/2p_0^{1/2}$, or solving for $p_0 = [(E_1/E_0) + ((E_1/E_0)^2 - 1)^{1/2}]^2$. These formulas show that the parameter indeed exists for any value of the $E_1$ energy.

From this initial example, it becomes clear that the parameter method is applicable to a variety of potential energy functions, for instance, a potential $V(r) = ar^b + c/r^d$, with all positive a, b, c, d, constants. In a similar way we get, $E_1/E_0 = b(1+p_0)(d/bp_0)^{d/(b+d)}/(b+d)$. Next, we assume that the attractive part $V_A(r)$ of the potential function is unknown, in order to see the scope of the parameter method. Introducing $V_A(r)$ in the total potential, we have

$$V(r) = V_A(r) + k_C qQ/r \tag{8}$$

and, proceeding in a similar way to Eq. (6), the equilibrium is given by the condition

$$V_A(r) = p_0 k_C qQ/r, \tag{9}$$

which has the solution for $r = r_p$ as

$$V_A(r_p) = p_0 k_C qQ/r_p. \tag{10}$$

Therefore the value of the total potential energy function (8) at equilibrium is

$$V(r_p) = (1 + p_0)k_C qQ/r_p, \tag{11}$$

which can be compared with the particular case of Eq. (4). According to our initial hypothesis, there are values of $p_0$ where the classical or quantum equilibrium is found. In more realistic situations, as in the last section, we have additional experimental data relating the values of the variables of the equilibrium that would help determine the parameter and other related quantities. Very similar steps from (8) to (11) will be useful again in the last section where the model of ionic crystals is presented.

In the next section, the parameter method is combined with an approximate method for quantum mechanics.

## III. THE KINETIC ENERGY AS A REPULSIVE POTENTIAL

It is possible to similarly apply the parameter method when there is only one attractive position potential. In this situation, the kinetic energy (KE) is treated like a repulsive potential. First, we consider a classic and then a quantum mechanical case. It is important to point out that, in this application, the value of the parameter in the following examples is not necessarily directly related to the conditions determining the quantum energy spectrum, as it was in the previous section. Here we show how the parameter method can be used to obtain a quick and valuable approximation to the energy spectrum without the use of the Schrödinger equation, which in most instances requires complex and difficult calculations. We begin with the total energy given by

$$E = KE + V = mv^2/2 + c_0 r^b, \qquad (12)$$

where r is the radial coordinate of the mass m, and v = speed of m. The coefficient and the exponent may be positive or negative with the condition

$$c_0 b > 0, \qquad (13)$$

so that, in this way, the radial potential energy is attractive. Applying the energy balancing parameter concept between the attractive and repulsive potentials, we have

$$|c_0| r^b = p_0 mv^2/2. \qquad (14)$$

The absolute value has been introduced to keep $p_0$ a positive parameter. Using this last Eq. in (12), the equilibrium total classical energy is

$$E = c_0 r^b (1 + \text{sgn}(c_0)/p_0), \qquad (15)$$

with $\quad \text{sgn}(c_0) = \text{sgn}(b) = +1 \text{ or } -1.$

The last condition on the sign is due to condition (13). If this is classical motion, Eq. (14) is a radial equation of motion in disguise; in such a case, if it is circular motion, the parameter is

$$p_0 = 2\text{sgn}(b)/b = 2/|b| \quad \text{(classical circular motion)}. \qquad (16)$$

Next, we consider quantum motion. We find a good estimate of the value of the quantum total energy using the arguments of Louis de Broglie (1924)[2]. His 1st argument is the introduction of quantum wave motion using the momentum of a mass particle to define the wavelength λ of the particle wave as follows:

$$mv = h/\lambda, \qquad (17)$$

where h = Plank's constant. Eq. (17) is now replaced in Eq. (14), where we find

$$|c_0|r^b = p_0 h^2/2m\lambda^2. \quad (18)$$

Now the 2$^{nd}$ de Broglie's argument is introduced. This is the presumption that the proposed matter waves also behave like "standing waves" when bounded. Louis de Broglie originally proposed his idea to provide a wave explanation to the Niels Bohr model of the atom (1913)[2,3]. This proposal was made in his Ph. D thesis (1924), undoubtedly one of the most extraordinary doctoral theses in the history of science. We have to remember that in those years was a radical idea to suggest that a material particle like an electron can behave as a wave. His thesis advisor, Paul Langevin, apparently unsure of what to do, showed the thesis to Albert Einstein, who approved it[4]. However, since the development of the Schrödinger wave equation (1926), de Broglie's implementation of the standing waves has been neglected.

Here we explore the idea that this 2$^{nd}$ argument is more valuable than previously considered. With a small generalization of the argument we obtain a quick approximation of the energy spectrum across many different energy potentials, something that would be very difficult to do by any other method. Louis de Broglie's 2$^{nd}$ argument uses the analogy to simple mechanical standing waves, implying that the matter wave can only have stable motion if the wavelength fits an integral number of times in the motion of the wave around its center,

$$n\lambda = 2\pi r p_1 \qquad n = 1, 2, 3, \ldots (\text{or a subset}), \quad (19)$$

where $p_1$ is a dimensionless parameter accounting for wave motion without circular symmetry, and $p_1 = 1$ for wave motion with circular symmetry (de Broglie's version). A subset of the natural numbers in (19) is also a possible condition found in simple standing waves, for example, the odd numbers are selected in the standing sound waves in a pipe closed at one end. Combining Eq. (18) with Eq. (19), we have

$$|c_0|r^b = p_0 h^2 n^2 / 8\pi^2 r^2 p_1^2 m. \quad (20)$$

This is an equation for the quantization of average orbital motion distance r, in terms of the n integer; then solving (20) for r, we find

$$r = r_n = (p_0 h^2 n^2 / 8\pi^2 m p_1^2 |c_0|)^{1/(b+2)}. \quad (21)$$

We replace this last equation in Eq. (15)

$$E = E_n = (1 + \text{sgn}(c_0)/p_0) c_0 (p_0 h^2 n^2 / 8\pi^2 |c_0| m p_1^2)^{b/(b+2)}. \quad (22)$$

Before analyzing (22) in detail, the next step will help us see more clearly the many physical results contained in (21) and (22). Every physical system, where a position potential energy is known, should have a characteristic length constant $r_0$. We can use $r_0$ to redefine $c_0$ in (12-15) in the following way:

$$c_0 = \text{sgn}(b) h^2 / 8\pi^2 m r_0^{(b+2)}. \quad (23)$$

Doing this change in (21) by substituting $c_0$, we find

$$r_n = r_0(p_0/p_1^2)^{1/(b+2)}n^{2/(b+2)}, \tag{24}$$

and in (22)

$$E_n = p_E \text{sgn}(b)(h^2/8\pi^2 m r_0^2)n^{2b/(b+2)}, \tag{25}$$

$$p_E \equiv (1 + (\text{sgn}(b)/p_0))(p_0/p_1^2)^{b/(b+2)}.$$

This last equation is now easily examined. It is the correct quantization of the energy for several values of the exponent b: the Coulomb's potential $b = -1$ (hydrogen-like atoms, and as L. de Broglie had obtained it); the free particle $b = 0$; and almost correct for the linear potential, $b = 1$; the harmonic potential $b = 2$; and the particle in a spherical box ($b \rightarrow \infty$ = "infinite").

The last cases ($b = 1, 2, ...$) of Eq. (25) require a more careful commentary. Here the spectrum of the harmonic potential in three dimensions is found exactly, if n are the odd numbers $n = 2k + 3 = 2(k + 3/2)$, $k = 1, 2, 3, ...$ . The particle in a spherical box of radius $r_0$ would require the limit $b \rightarrow +\infty$ and, in this limit, $p_0$ is kept constant. Also, the spectrum in a spherical box is proportional to $n^2$ only for large n, or for zero angular momentum ($L = 0$), when compared to the exact results (Schrödinger's Eq.). The linear potential ($b = 1$) has an energy spectrum proportional to $n^{2/3}$ also for large n or $L = 0$. The case $b = -2$ is a exponent singularity; however, tracing it back to Eq. (20), this is a case where there is not quantization of the average orbital distance. It seems that, when $b \approx -2$, the results of Eq. (22) are not valid. Eq. (25) provides a notable interpolation for the energy spectrum of a variety of potentials. Many of those potentials, such as b = fractional numbers, are considered analytically unsolvable problems when treated by means of the Schrödinger equation.

Also noteworthy, are the negative values of $b < -2$. For very large negative values of b ($b \rightarrow -\infty$), the energy spectrum is proportional to $-n^2$, similar to the particle in a box of radius $r_0$ but without bottom for the energy, that is, no finite ground state value. For example, from (25), if $b = -6$, the potential energy is proportional to $-(r_0/r)^6$ and the energy levels are proportional to $-n^3$. This bottomless spectrum is a general characteristic when $b < -2$ in this model. Are the potentials with $b < -2$ unphysical due to this last characteristic? This is an important point requiring additional research.

Looking back to Eq. (24), we have a direct interpretation of the parameter $p_0$, which is given as the ratio of the first average quantum length $r_1$, and $r_0$

$$p_0 = p_1^2(r_1/r_0)^{b+2}. \tag{26}$$

In formal quantum mechanics, energy quantization is not a result of the average orbital distance quantization. However, the wave functions ($\psi_n$) of the energy spectrum ($H\psi_n = E_n\psi_n$) have an average orbital distance value ($\int |\psi_n|^2 r dV$) that is a discrete sequence similar to Eq. (24). The 1st de Broglie argument is believed to be exact and, in fact, can be used to make a heuristic derivation of the Schrödinger equation ($H\psi = E\psi$). This step is shown in the Appendix of this article. The 2$^{nd}$ de Broglie argument as presented in this section, is a good approximation to some aspects of quantum mechanics.

## IV. BINDING ENERGY OF IONIC CRYSTALS

Ionic crystals are salts similar to NaCl, and also known as ionic compounds. This section shows the application of the parameter method to the Born & Lande[3] model (1918) of the ionic crystals bond energy. This is a well known topic of physics, currently found as a chapter of physical chemistry[3].
The modeling of the bond energy of the crystal is based on the positive and negative ions attracting each other with the usual electrostatic potential ($-1/r$) to a point where a repulsive potential ($V_R$) takes over, thus reaching equilibrium. Several forms have been suggested for the repulsive potential, one of them $V_R \propto 1/r^n$, n >1. Although the repulsive potential is not well known, a result for the energy of the crystal is attainable.
The potential energy of all ions of the crystal is

$$V = (1/2)\Sigma_{k,j}[k_C Q_j Q_k/r_{kj} + V_R(r_{kj})] \tag{27}$$

where $Q_k$ are the ionic charges (with their + and – signs), $k_C$ = Coulomb's constant, and $r_{kj}$ = distance between j and k ions. The (1/2) factor is due to the fact that, if the summations are done for all pair combinations for each ion, then every interactive pair is counted twice. There are only one type of positive and one type of negative ions in the crystal. The model also assumes that most of the energy of the crystal is due to this potential energy, implying that the kinetic energy is small. If the kinetic energy is nearly zero, then this is the energy of the crystal for absolute temperature = T ≈ 0. The next step introduces the distance $r_0$, which is the nearest-neighbor ion separation distance, that is used to redefine the distances as

$$r_{kj} = d_{kj}r_0. \tag{28}$$

Replacing this last into (27)

$$V = (1/2)\Sigma_{k,j}[Q_j Q_k/(e^2 d_{kj})](k_C e^2/r_0) + (1/2)\Sigma_{k,j} V_R(d_{kj}r_0) \tag{29}$$

$$= N[-C_M k_C e^2/r_0 + V_{ER}(r_0)],$$

where the parameter $C_M = -\Sigma_k[Q_1 Q_k/(e^2 d_{1k})]$, is the Madelung's dimensionless constant, and N is the number of ion pairs. The repulsive potential also becomes an effective function $V_{ER}(r_0)$. If the repulsive potential is a power function ($V_R \propto 1/r^n$), then it can be reduced in the same way as the electric potential. However, as we shall see, the equivalent to the Madelung's constant for the repulsive potential is not necessary. Applying the parameter method discussed in the previous sections, we have that the equilibrium is found for a value of the parameter $p_0$ balancing the repulsive and attractive potentials

$$V_{ER}(r_0) = p_0 C_M k_C e^2 / r_0. \tag{30}$$

Introducing (30) in (29), we have

$$V = -(1 - p_0) N C_M k_C e^2 / r_0, \tag{31}$$

where it is clear now that $0 < p_0 < 1$ in this case. Eq. (31) is the classical formula of Born & Lande (1918) for the energy of ionic crystals[3]. The Madelung's constant can be calculated separately using the crystallographic information on the $r_{kj}/r_0$ distances, and generally it is not a simple task. Experimental data shows that $p_0 \approx 0.1$ depending on the type of crystal.

In conclusion, the value of the parameter $p_0$ is a combination of classical and quantum effects, as it was described in the 2$^{nd}$ section of this article. Surprisingly, in the end, we do not need the repulsive potential in exact detail; instead, the value of the parameter $p_0$ is sufficient to determine the bond energy of the crystal.

**ACKNOWLEDGMENTS**
I would like to thank M.-C. Vuille for her assistance in editing and helping make this a more readable article.

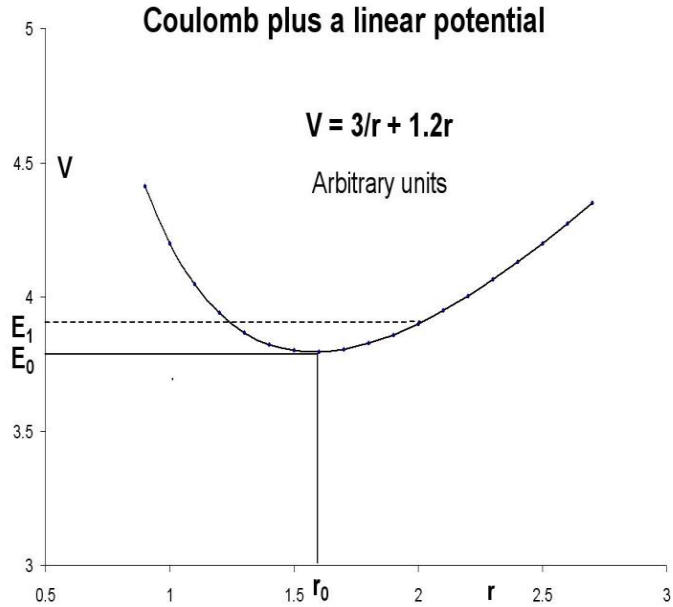

Fig. 1. Potential energy showing the lowest classical energy $E_0$ and the lowest quantum energy $E_1$, see Eq. (5).

# APPENDIX: ON THE SCHRÖDINGER EQUATION HEURISTICS

The standard method of deriving the Schrödinger equation and most of the important equations of quantum mechanics relies on the operator formalism and the canonical association of the operators with specific physical quantities. This operator formalism is effectively a new set of physical principles not derivable from any other area of physics, though many classical concepts and formulas are an integral part of these operator principles. This quantum mechanical formalism is very important, but unrevealing on the wave nature of the solutions of the equations. In my opinion there is an intuition gap here because the fundamental equation of the energy spectrum, with its associated eigenvalue functions (the Schrödinger equation), is not at a first glance an obvious type of wave equation with solutions similar to standing waves. Standard treatments of the possible wave solutions show that the free waves are indeed solutions of the equation or solve very simple one-dimensional examples. However, these treatments are still short of bridging the intuition gap, especially since the profound impact of the Schrödinger equation is in the realistic bounded three-dimensional particles and interacting particles, and not in the free particles.

By adapting the classical wave equation and with the help of the 1$^{st}$ de Broglie argument, it is possible to obtain the time-independent Schrödinger equation. This derivation has been known for a while; I saw it for the first time as an exercise in the textbook of Bromberg[3], and it is probably older than that. In this way even without solving any Schrödinger equation, it is evident that we are dealing with some type of wave equation. A more important place should be given to this derivation.

Starting with the classical wave equation, a wave $\Psi$ moving with speed $v = \lambda f =$ (wavelength)(frequency), satisfies

$$\Delta\Psi = (1/v^2)\partial^2\Psi/\partial t^2; \quad \Delta \equiv \partial^2/\partial x^2 + \partial^2/\partial y^2 + \partial^2/\partial z^2 \quad\quad (A1)$$

and, assuming the following time dependence for a wave,

$$\Psi = \Psi(x,y,z)\sin(2\pi ft + \phi) \quad\quad (A2)$$

we get from (A1)

$$\Delta\Psi(x,y,z) = (-4\pi^2 f^2/v^2)\Psi(x,y,z), \quad\quad (A3)$$

which effectively can be taken as an equation of the space portion of the waves.

Next, we introduce the 1$^{st}$ de Broglie's argument, the momentum p of the particle being inversely proportional to the wavelength of the particle,

$$p = h/\lambda = hf/v, \quad\quad (A4)$$

with the adaptation that the momentum is also space dependent, through the momentum formulation of the total energy E

$$E = p^2/2m + V(x,y,z). \quad (A5)$$

Combining the last two equations,

$$p^2 = 2m(E - V(x,y,z)) = (hf/v)^2 \quad (A6)$$

which, once substituted in (A3), becomes the Schrödinger time independent equation

$$[- ((h/2\pi)^2/2m)\Delta + V]\Psi(x,y,z) = E\Psi(x,y,z). \quad (A7)$$

Finally, there is the important question: what is waving in (A7)? A persistent question, since Schrödinger's first proposal. There is a working answer in standard quantum mechanics, but apparently Schrödinger never approved it. He was inclined to the idea that the wave function represents actual motion of the electric charge density. For a review of the historical context of this and other enduring controversies in quantum mechanics, see Bacciagaluppi[5].

In the classical waves, $\Psi(x,y,z)$ is either a displacement of a macroscopic amount of matter (more than thousands of atoms) or a field with a macroscopic amount of energy or wavelength. The wave function in (A7) cannot have the same classical interpretation, since this wave function in principle applies to less than one atom, such as a subatomic particle. The accepted answer to the question of what is waving in (A7) is that the square of the absolute value of the wave functions ($\Psi$ may be a complex value function) is equal to the probability density of finding the particles, that is, probability density = $|\Psi|^2$. This last answer remains a difficult chapter of quantum mechanics, where heuristics does not yet reach. However, the Paul Ehrenfest's theorem provides the closest interpretation of the quantum wave function in connection with classical concepts, and support for the probabilistic interpretation of the quantum wave function.

Dear Editors:

Herewith I submit to for consideration towards publication my article "The Energy Balancing Parameter". The article contains one Appendix and one Figure.

This article presents a method suitable for the analysis of the energy of a system where a full knowledge of the potential energy is not available. Also, this method can be applied to obtain a good approximation in a quantum system.

The parameter method is intuitive in the semi-classical sense and mathematically simple, so much so that I think an undergraduate student with interest in the subject matter, having only basic training in algebra and calculus, can easily follow the arguments presented.

The results are compelling. Good estimates of the energy spectrum are obtained for a variety of physical systems, without using differential equations (Schrödinger's equation), and without having to explain the intricate quantum operator formalism. To show the wide range of potential uses of the parameter method, it is applied to ionic crystals, and a successful derivation of the standard Born & Lande model is made.

Thank you for your time and attention.

Sincerely,

Walton R. Gutierrez, Ph. D.
Touro College
27 W. 23rd St.
New York, N Y 10010

E-mail: waltong@touro.edu